\documentclass[amsmath,amssymb,prc,final,showpacs,twocolumn]{revtex4-1}
 
\usepackage{bm,hyphenat,xspace}
\usepackage{graphicx,epsfig}

\newcommand {\mbf}[1]{{\mathbf{#1}}}

\newcommand {\mcu}{\mathcal{U}}

\newcommand {\mcj}{\mathcal{J}}
\newcommand {\mct}{\mathcal{T}}

\begin{document}

\title {Universality in fermionic dimer-dimer scattering}
  
\author{A.~Deltuva} 
\email{arnoldas.deltuva@tfai.vu.lt}
\affiliation
{Institute of Theoretical Physics and Astronomy, 
Vilnius University, Saul\.etekio al. 3, LT-10257 Vilnius, Lithuania
}

\received{\today} 

\begin{abstract}
Collisions of two fermionic dimers near the unitary limit are studied
using exact four-particle equations for transition operators
in momentum space.  Universal properties of dimer-dimer
phase shifts and effective range expansion (ERE) parameters are determined. 
The inclusion of the fourth-order momentum term in the ERE significantly
extends its validity to higher collision energies.
The dimer-dimer scattering length and effective range
are determined in the unitary limit as well as their corrections
arising due to the finite range of the two-fermion interaction.
These results are of considerably higher accuracy as compared 
to previous works, but  confirm most of the previous results
except for the lattice effective field theory  calculations.
\end{abstract}

 \maketitle

\section{Introduction \label{sec:intro}}

Few-particle systems with large $s$-wave scattering length $a$ exhibit
universal properties independent of the short-range interaction details.
One of the most prominent examples is the Efimov effect \cite{efimov:plb},
the emergence of an infinite number of weakly bound three-particle states
(trimers) with total orbital angular momentum $L=0$
and geometric energy spectrum in the unitary limit $a \to \infty$.
Furthermore, there are four-particle (unstable) bound states 
(tetramers) associated with each trimer 
\cite{hammer:07a,stecher:09a,deltuva:10c,deltuva:11a}.
However, the Efimov effect is prohibited in the system of identical
spin $\frac12$ fermions (including both components, i.e., spin
up and down states), due to the antisymmetry of the wave function.
The only bound state in such a system with large $a$
(assuming the absence of deeply  bound states) 
is a weakly-bound dimer formed by two fermions with antiparallel spins;
absolute value of its binding energy is approximately given by
$\varepsilon_2  \approx \hbar^2/ma^2$ where $m$ is the fermion mass 
(unit convention $\hbar = 1$ is adopted in the present work).
The total spin, orbital and total angular momentum of the dimer are zero,
i.e., in the standard spectroscopic ${}^{2S+1}L_J$ notation it is realized
in the  ${}^{1}S_0$ partial wave. Although there are no four-fermion
bound states, the four-body physics is important for the properties
of cold dilute molecular gases that are determined by the parameters
of low-energy dimer-dimer collisions
\cite{petrov:04a,PhysRevA.77.043619,PhysRevA.79.030501}.
The dimer-dimer scattering length $a_{dd}$ has been investigated
in a number of works 
\cite{petrov:04a,PhysRevA.77.043619,PhysRevA.79.030501,bulgac:03a,elhatisari:17a},
all predicting $a_{dd}/a \approx 0.6$ in the large $a$ limit,
but the studies at finite collision energies are  scarcer and contradicting
\cite{PhysRevA.77.043619,PhysRevA.79.030501,elhatisari:17a}.
 The correlated Gaussian (CG) \cite{PhysRevA.77.043619},
the fixed-node diffusion Monte Carlo (FN-DMC) \cite{PhysRevA.77.043619}, and
the hyperspherical adiabatic (HA) \cite{PhysRevA.79.030501}  approaches 
 predict the  dimer-dimer effective range $r_{dd}/a$  to be
 $0.13(2)$, $0.12(4)$, and  $0.13$, respectively,
while  the recent lattice effective field theory (L-EFT) calculation 
\cite{elhatisari:17a}
with thorough systematic error estimation claims a very different result
$r_{dd}/a = -0.431(48)$.
One has to admit that most of the above approaches are based on 
a finite-volume approximation and need extrapolation to get free-space
results. This is not needed for the exact formulation
of  the four-particle scattering problem as proposed by 
Faddeev and Yakubovsky \cite{yakubovsky:67} and by
Alt, Grassberger, and Sandhas (AGS) \cite{grassberger:67}.
In the latter case the properly symmetrized equations for the 
four-particle transition operators have been applied to the study of 
the four-boson Efimov physics 
\cite{deltuva:10c,deltuva:11a,deltuva:11b}.
The AGS equations were solved numerically in the momentum-space
 partial-wave representation leading to the most accurate results
for the bosonic particle-trimer and dimer-dimer scattering processes.
The present works aims to extend the methodology of
 Refs.~\cite{deltuva:10c,deltuva:11a,deltuva:11b}
for the scattering of two fermionic dimers and to provide accurate 
benchmark results.

Section \ref{sec:eq} contains dimer-dimer scattering equations
and some details of calculations,
results are reported in Sec. III,
 and conclusions are presented in Sec. IV.

\section{Theory \label{sec:eq}}

The AGS equations \cite{grassberger:67} in the symmetrized form suitable
 for the bosonic dimer-dimer 
scattering were given in  Ref.~\cite{deltuva:11b}. Using the spin formalism,
spin up and down fermions can be considered as different states
of identical particles, thereby allowing to consider the system
of four identical fermions and antisymmetrize the AGS equations
in a corresponding way.
 Thus, the scattering equations
 can be generalized to include formally both bosonic and fermionic systems 
by introducing the symmetry parameter $\zeta$ being $+1$ for bosons and
$-1$ for fermions, i.e.,
\begin{subequations} \label{eq:U}
\begin{align}  
\mcu_{12}  = {}&  (G_0  t  G_0)^{-1}  
 +\zeta P_{34}  U_1 G_0  t G_0  \mcu_{12} + U_2 G_0  t G_0  \mcu_{22} , 
\label{eq:U12} \\  
\mcu_{22}  = {}& (1 + \zeta P_{34}) U_1 G_0  t  G_0  \mcu_{12} . \label{eq:U22}
\end{align}
\end{subequations}
The four-particle  transition operators $\mcu_{\beta\alpha}$ 
 are labeled according to the two cluster partitions,
$\beta =1$  standing for the 3+1  partition  (12,3)4, 
and $\beta =2$  standing for the 2+2  partition  (12)(34).
The two-particle transition matrix 
\begin{equation} \label{eq:t}
t= v + vG_0 t
\end{equation}
 is calculated
from the potential $v$ where $G_0 = (E+i0-H_0)^{-1}$ is the free 
four-particle resolvent at the available 
energy $E$ in the center-of-mass (c.m.) frame and $H_0$ is the
free Hamiltonian for the relative motion. Furthermore,
\begin{equation} \label{eq:AGSsub}
U_\beta =  P_\beta G_0^{-1} + P_\beta t\, G_0 \, U_\beta
\end{equation}
are the
transition operators for  the 1+3 and 2+2 subsystems with
$P_1 =  P_{12}\, P_{23} + P_{13}\, P_{23}$, $P_2 =  P_{13}\, P_{24}$ and
the permutation operators $P_{ab}$ of particles $a$ and $b$.

The AGS equations (\ref{eq:U}) are solved in the momentum-space
partial-wave framework where they become a system of integral equations 
with three continuous variables, the magnitudes of the 
 Jacobi momenta  $k_x , \, k_y$ and $k_z$ \cite{deltuva:12a};
 the associated orbital angular momenta are $l_x$, $l_y$, and $l_z$.
As compared to the bosonic spin zero case \cite{deltuva:10c,deltuva:11a,deltuva:11b,deltuva:12a},
the basis states have to be extended to include the spins $s_i = \frac12$.
The states of the  total angular momentum  $\mathcal{J}$ 
with projection  $\mathcal{M}$ are 
$ | k_x \, k_y \, k_z   
\{l_z [(l_y \{[l_x (s_1s_2)s_x]j_x \, s_3\}S_y ) J_y s_4 ] S_z \} 
\,\mathcal{JM} \rangle$ 
for the 3+1 configuration and 
$|k_x \, k_y \, k_z  (l_z  \{ [l_x (s_1s_2)s_x]j_x\, [l_y (s_3 s_4)s_y] j_y \} S_z)
\mathcal{ J M} \rangle $ for the 2+2 configuration;
 the calculation of $U_\beta$ is performed in the corresponding basis
while transformations from one basis to another are needed 
in certain steps of the solution process  \cite{deltuva:07a}.
 The discrete quantum numbers $j_x$ and $j_y$ are the total angular 
momenta of pairs (12) and (34),
$J_y$ is the total angular momentum of the (123) subsystem,
and $s_x$,  $s_y$, $S_y$, and $S_z$ are the intermediate subsystem spins.
To  ensure the full antisymmetry of the four-fermion system,
the basis states must
 be antisymmetric under exchange of two fermions in the 
subsystem (12) for the $3+1$ partition 
and in (12) and (34) for the $2+2$ partition, i.e.,
$l_x + s_x$ (and $l_y + s_y$ for the $2+2$ configuration) must be even.
When solving equations (\ref{eq:U}) numerically, the integrals
are discretized using Gaussian quadratures. The absence of 
near-threshold $\mcu_{\beta\alpha}$  poles allows for an accurate solution
using iterative methods such as the double Pad\'e summation 
of  Ref.~\cite{deltuva:07a}. Thus, there is no necessity for the
direct matrix inversion and separable form of $v$, $t$,  $U_\beta$,
and  $\mcu_{\beta\alpha}$ as used in the four-boson calculations
\cite{deltuva:10c,deltuva:11a,deltuva:11b,deltuva:12a}.
Below the dimer breakup threshold
the only singularity in the kernel arises from the dimer-dimer pole in $U_2$; 
it is treated by the  subtraction method \cite{deltuva:07a}.
The partial-wave amplitude for the elastic dimer-dimer scattering 
with the relative on-shell momentum $p_{dd}$ and
 $E = -2\varepsilon_2 + p_{dd}^2/2m$ is obtained as
\begin{equation} \label{eq:F}
 \mct^{\mcj}_{dd}(p_{dd}) =  
2 \langle \phi_2^{\mcj}(p_{dd}) |\mcu_{22} |\phi_2^{\mcj}(p_{dd}) \rangle 
\end{equation}
where
$|\phi_2^{\mcj}(p_{dd}) \rangle = G_0 \, t  P_2 |\phi_2^{\mcj}(p_{dd}) \rangle$ 
is the $\mcj$-component
of the Faddeev amplitude of the asymptotic channel state 
$|\Phi_2(\mbf{p}_{dd}) \rangle = (1+  P_2) |\phi_2(\mbf{p}_{dd}) \rangle$.
Given the symmetry restrictions, only even $\mcj$ contribute to the
dimer-dimer scattering; furthermore, for each  $\mcj$ the 
Faddeev amplitude $|\phi_2^{\mcj}(p_{dd}) \rangle$ has a single component
with $l_x=s_x=j_x=l_y=s_y=j_y=S_z =0$ and $l_z = \mcj$.
The scattering amplitude (\ref{eq:F}) leads to the single-channel 
$S$-matrix and  phase shift $\delta_{\mcj}$ as
\begin{equation} \label{eq:S}
 \mathcal{S}^{\mcj}_{dd}(p_{dd}) =  e^{2i\delta_{\mcj}} = 
1 - 2i \pi m p_{dd} \mct^{\mcj}_{dd}(p_{dd}).
\end{equation}
Further details of the numerical methods used to solve the AGS equations can be found in 
Ref.~\cite{deltuva:07a}.

\section{Results \label{sec:res}}

To study the universality in fermionic dimer-dimer collisions 
and to prove the independence of the short-range interaction details,
three types of potential models are used in the present work;
all of them are assumed to act in the $s$-wave $l_x=0$ only.
The first one is a separable potential
$v(k_x',k_x) = g(k_x') \lambda g(k_x) $
 with Gaussian form factors  $g(k_x) = e^{-(k_x/\Lambda)^2} $.
The cutoff parameter $\Lambda$ and the strength  $\lambda$
are adjusted to reproduce the desired values of 
two-fermion scattering length $a$ and effecttive range $r_e$.
The second type corresponds to a fictitious four-neutron system
with enhanced two-neutron interaction supporting a weakly
bound dineutron. For this purpose a realistic charge-dependent
 Bonn (CD Bonn) potential \cite{machleidt:01a} multiplied by a factor ranging from
1.1 to 1.17 is used; such a variation is sufficient to control
the dimensionless ratio  $r_e/a$ that characterizes the deviation
from the unitary limit. 
In a similar way, the third type corresponds to a 
fictitious system of ${}^3\mathrm{He}$ atoms interacting through the 
modified  LM2M2 potential \cite{aziz} multiplied by a factor around 1.3;
the transformation to the momentum space is performed as described in 
Ref.~\cite{deltuva:15g} taking $r_{\rm min} = 2.3$ fm and 
 $r_{\rm max} = 50$ fm.  In the following
the three potential types will be labeled Vsep, CDBx, and LM2M2x,
respectively. Since all models differ also in 
$\varepsilon_2$, the collision energy will be characterized by 
another dimensionless quantity $ap_{dd}$; the dimer breakup threshold
corresponds to $ap_{dd} \approx \sqrt{2}$.

\begin{figure}[!]
\begin{center}
\includegraphics[scale=0.66]{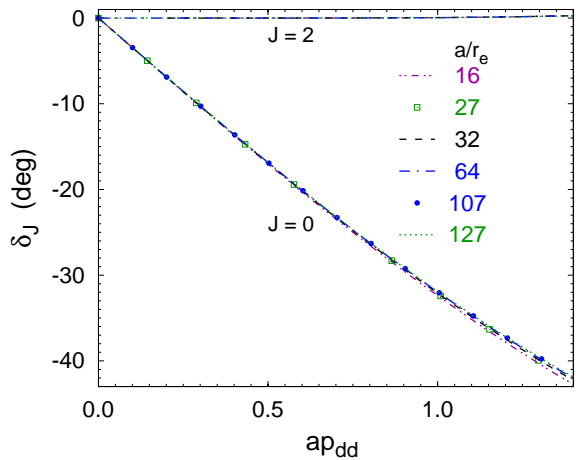}
\end{center}
\caption{\label{fig:phase} (Color online)
Phase shifts for fermionic dimer-dimer scattering in $s$ and $d$ waves.
Results are obtained for different ratios $a/r_e$ using the models Vsep (curves),
 CDBx (full circles), and LM2M2x (boxes). }
\end{figure}

The calculated phase shifts  $\delta_0$ and  $\delta_2$  are presented in
Fig.~\ref{fig:phase}. The latter remains very small over the whole
 energy regime; thus, the dimer-dimer scattering is 
 strongly dominated by the relative $s$-wave. The results are shown 
 for six different  $a/r_e$ values ranging from 16 to 127; only the curves
for lower $a/r_e$ are discernible, indicating that higher $a/r_e$ predictions
approximate well the unitary limit results. 

\begin{figure}[!]
\begin{center}
\includegraphics[scale=0.64]{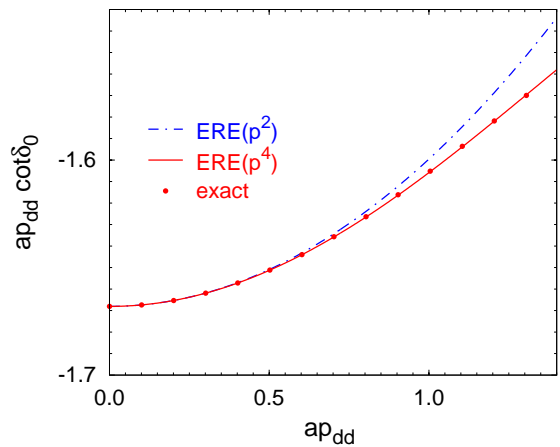}
\end{center}
\caption{\label{fig:ere24} (Color online)
Function  $ap_{dd}\cot{\delta_0}$, calculated directly at  $a/r_e \approx 127$,
is compared with its ERE with/without the fourth-order momentum term. }
\end{figure}

Is is important to evaluate also the
theoretical error bars. They are dominated by the truncation of the partial-wave
expansion for the 3+1 configuration and by the numerical accuracy 
of integrations and interpolations
when solving the scattering equations. The results  in
Fig.~\ref{fig:phase} include orbital angular momenta $l_y$ and $l_z$ up to 4,
i.e., $J_y$ up to $\frac92$. Reducing to $l_y,l_z \leq 3$, $J_y \leq \frac72$ 
changes of $\delta_0$ by 0.005 deg at most, while increasing to 
 $l_y,l_z \leq 5$, $J_y \leq \frac{11}{2}$ leads to changes by 0.001 deg or less. 
The number of momentum discretization grid points increases from
30 (for lowest  $a/r_e$) to 70 (for highest  $a/r_e$); the  $\delta_{0}$ predictions
are stable within 0.001 to 0.005 deg for reasonable variations of grid
point distributions. Due to the presence of significant high-momentum
components, the accuracy is  lower in  LM2M2x calculations.
Otherwise, the theoretical error bars for the phase shift results 
 are  well below 0.01 deg, enabling  an accurate extraction of the
effective range expansion (ERE) parameters. To extend the validity of the ERE
to higher collision energies,
fourth-order momentum term is included, i.e.,
\begin{equation} \label{eq:kctg}
ap_{dd}\cot{\delta_0} \approx -\frac{a}{a_{dd}} + \frac12 \frac{r_{dd}}{a}
(ap_{dd})^2 - \frac14 c_{dd} (ap_{dd})^4.
\end{equation}
The importance of this term is illustrated in Fig.~\ref{fig:ere24}, comparing
directly calculated $ap_{dd}\cot{\delta_0}$ and its ERE. The second-order ERE
at $ap_{dd} > 0.5$ deviates from the exact results while the fourth-order ERE fits
the exact results very well over the considered energy regime $ap_{dd} < 1.4$.

\begin{figure}[!]
\begin{center}
\includegraphics[scale=0.64]{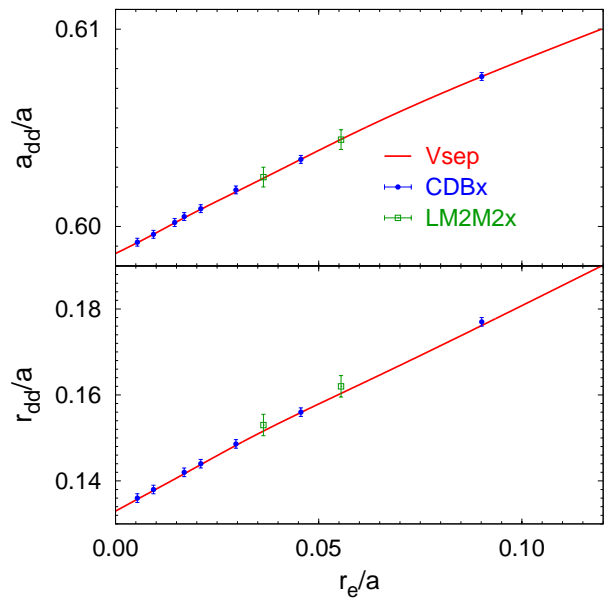}
\end{center}
\caption{\label{fig:ar} (Color online)
Dimer-dimer scattering length and effective range near the 
unitary limit calculated using Vsep (curves), CDBx (full circles),
and LM2M2x (boxes) models. }
\end{figure}

\begin{figure}[!]
\begin{center}
\includegraphics[scale=0.64]{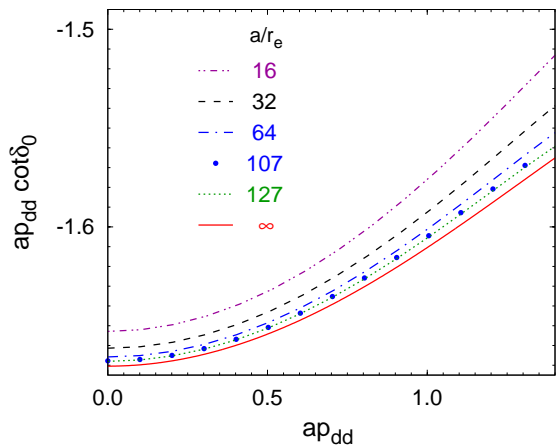}
\end{center}
\caption{\label{fig:ere} (Color online)
Function  $ap_{dd}\cot{\delta_0}$ for finite  $a/r_e$ values
is compared with its unitary limit (solid curve) calculated using parameters  
from Eqs.~(\ref{eq:ar}). }
\end{figure}

The convergence of the dimensionless ERE parameters $a_{dd}/a$ and $r_{dd}/a$
towards the unitary limit is studied in Fig.~\ref{fig:ar}.
The predictions for a number of Vsep (curves), CDBx (full circles),
and LM2M2x (boxes) models 
are plotted as functions of the respective $r_e/a$ values. The CDBx results 
include the  uncertanties resulting from the ERE fit procedure. They are
0.0002 for  $a_{dd}/a$ and 0.001 for  $r_{dd}/a$.
The error bars are the same for Vsep predictions but are not shown,
while LM2M2x predictions have larger error bars.
Within given error bars the agreement between Vsep, CDBx, and LM2M2x results
is perfect, indicating the independence of the short range interaction details.
 $a_{dd}/a$ and  $r_{dd}/a$ to a good accuracy
correlate with $r_e/a$ linearly at  $r_e/a < 0.04$, enabling not only a
 reliable extrapolation
to the unitary limit but also a systematic evaluation of finite range effects. 
Taking into account  Vsep and CDBx results with 
the error bar estimation from all uncertainty sources, the relations are 
\begin{subequations} \label{eq:ar}
\begin{align}  
\frac{a_{dd}}{a} =  {}& 0.5986 + 0.105\frac{r_{e}}{a} \pm 0.0005, \label{eq:ara} \\  
\frac{r_{dd}}{a} =  {}& 0.133  + 0.51 \frac{r_{e}}{a} \pm 0.002. \label{eq:arr} 
\end{align}
The fourth-order ERE coefficient $c_{dd}$ is small and has relatively large uncertainty,
i.e.,
\begin{gather}
{c_{dd}} =   0.026  - 0.1 \frac{r_{e}}{a} \pm 0.004. \label{eq:arc} 
\end{gather}
\end{subequations} 
On a relative scale, $r_{dd}$ and $c_{dd}$ exhibit substantially stronger $r_e$
dependence than  $a_{dd}$, i.e., as one could expect, 
the importance of finite range effects increases with increasing collision energy.
This can be seen clearly also in Fig.~\ref{fig:ere}, comparing
 $ap_{dd}\cot{\delta_0}$ results for a number models with finite 
$a/r_e$ and the unitary limit ERE: the spread of predictions becomes
wider with increasing  $ap_{dd}$.

\section{Summary \label{sec:sum}}

Fermionic dimer-dimer scattering near the unitary limit was studied
using exact four-particle equations for transition operators
that were solved numerically in the momentum-space partial-wave 
framework. Three types of interaction models were used proving
the independence of the results  of the short-range potential details
and thereby establishing universal behaviour of dimer-dimer
phase shifts and ERE parameters. 
The usual ERE up to the second-order in momentum was found to be valid for low
collision energies only, but including the fourth-order term
$ \frac14 c_{dd} (ap_{dd})^4$  as in 
Eq.~(\ref{eq:kctg}) extends the ERE validity up to the dimer breakup threshold.
The finite range $r_e$ of the two-fermion interaction leads to 
corrections for ${a_{dd}}/{a}$, ${r_{dd}}/{a}$, and  $c_{dd}$ that,
sufficiently close to the unitary limit,
are linear in $r_e/a$ as given in  Eqs.~(\ref{eq:ar}); furthermore,
the finite range correction increases with increasing collision energy.

The present results are of
considerably higher accuracy as compared to previous works.
All of them are consistent with the present value of the 
dimer-dimer scattering length in the unitary limit,
${a_{dd}}/{a} = 0.5986  \pm 0.0005$. The obtained effective range parameter
${r_{dd}}/{a} = 0.133  \pm 0.002$ supports previous 
CG, FN-DMC, and HA calculations of
 Refs.~\cite{PhysRevA.77.043619,PhysRevA.79.030501}
and indicates heavy failure of the L-EFT method \cite{elhatisari:17a}.
Since the lattice-type methods become used quite often, 
especially in nuclear physics,
it is very important to evaluate their reliability for scattering calculations.

Regarding the nuclear physics, the present work considers a fictitious 
four-neutron system with bound dineutrons. Nevertheless, this is an 
important step towards the study of  tetraneutron states
using realistic interactions.

\vspace{1mm}

The author acknowledges the support  by the Alexander von Humboldt-Stiftung and
the hospitality of the Ruhr-Universit\"at Bochum
where a part of this work was performed.


\end{document}